\documentclass[twocolumn,showpacs,pra,letter]{revtex4}%
\usepackage{amsmath,amssymb,amsthm}
\usepackage{epsfig}
\usepackage{epstopdf}
\usepackage{graphicx}%
\usepackage{amsmath}%
\usepackage{amsfonts}%
\usepackage{amssymb}

\begin{document}

\title{Speeding up Entanglement Degradation}
\author{R.B. Mann$\dagger$ $\ddag$ and V. M. Villalba$^{\dagger}$\footnote{On sabbatical leave from
the Center of Physics, IVIC, Apdo 21827 Caracas  1020A Venezuela} }
\affiliation{$\dagger$ Department of Physics \& Astronomy, University of Waterloo,
Waterloo, Ontario Canada N2L 3G1}
\affiliation{$^{\ddag}$Perimeter Institute, 31 Caroline Street North Waterloo, Ontario
Canada N2L 2Y5}

\begin{abstract}
We present a method for
tracking time-dependent entanglement between different modes of a quantum system
as measured by observers in different states of relative (non-uniform) motion.   By describing states on
a given spacelike hypersurface, observers/detectors in different states of motion can detect
different modes of excitation in a quantum field at any desired instant and thereby track various measures
of entanglement as function of time. We illustrate our method for a scalar field, showing how
entanglement degrades as a function of time if one observer begins in a state of inertial motion but ends in a state of uniform acceleration while the other remains inertial.

\end{abstract}
\pacs{03.65. -w , 03.65.Yz, 03.67.Mn}

\maketitle

\section{Introduction}

Entanglement is a key resource in quantum computational tasks \cite{ekertbook}
such as teleportation\cite{teleportation}, communication, quantum control
\cite{control} and quantum simulations \cite{simulations}. It is a property of
multipartite quantum states that arises from the superposition principle and
the tensor product structure of Hilbert space. It can be quantified uniquely
for non-relativistic bipartite pure states by the von Neumann entropy, and
several measures such as entanglement cost, distillable entanglement and
logarithmic negativity have been proposed for mixed states \cite{entanglement}.

Understanding entanglement in a relativistic setting is important both for
providing a more complete framework for theoretical considerations, and for practical
situations such as the implementation of quantum computational
tasks performed by observers in arbitrary relative motion. For observers in
uniform relative motion the total amount of entanglement is the same in all
inertial frames \cite{inertial}, although different inertial observers
may see these correlations distributed differently amongst various degrees of
freedom. However for observers in relative uniform acceleration a
communication horizon appears, limiting access to information about the whole
of spacetime, resulting in a degradation of entanglement as demonstrated for scalars
 \cite{ivette,ball} and fermions \cite{alsingspin}, and restricting the fidelity of  processes such as teleportation \cite{teleport},
and other communication protocols \cite{bradler}.

The most general situation is that of observers in different states of non-uniform motion.
This situation is relevant even for inertial observers in curved spacetime, who will undergo
relative non-uniform acceleration due to the geodesic deviation equation, and
therefore disagree on the degree of entanglement in a given bipartite quantum state.
While it is expected such entanglement will be time-dependent, there is presently no approach for
determining how the entanglement changes throughout the course of the motion.

We address this issue by constructing a new method for determining such time-dependent entanglement measures.
By considering the description of the field modes on a given spacelike hypersurface we are able
to compute the Bogoliubov transformations between states of instantaneous positive and negative
frequency as determined by observers in different states of non-uniform motion.  We illustrate our
approach by considering specifically the entanglement between two modes of a
non-interacting  scalar field when one of the observers describing the
state begins in a state of inertial motion and ends in a state of uniform
acceleration. From the perspective of the inertial observer (Alice) the state
is a maximally entangled pure state. However the non-uniformly accelerated
(NUA) observer (Vic) finds that the entanglement degrades as a function of
time, approaching at late times the constant value measured by  an observer
(Rob) undergoing the same asymptotic uniform acceleration. The paper is structured as follows: In Sec II we 
solve the Klein Gordon equation in the  NUA coordinates and compute the Bogoliubov coefficients. In Sec. III,  we quantify the
degree of entanglement with the help of the logarithmic negativity N and mutual information I. Finally, we summarize our results
in the conclusions

\section{Solution of the Klein Gordon equation}

For the inertial observer Minkowski coordinates $(t,x)$ are the most suitable
for describing the field, whereas for a uniformly accelerated (UA) observer Rindler
coordinates $(\tau,\xi)$ are appropriate. As these coordinates cover only a
quadrant of Minkowski space, the UA observer remains constrained to a particular
Rindler quadrant and has no access to the other Rindler sector. Both the inertial and UA observers
can define a vacuum state and a Fock space for a scalar field obeying the
Klein-Gordon equation, with solutions in each related by Bogoliubov transformations. A given
mode seen by an inertial observer corresponds for the UA observer to a two-mode squeezed state,  associated with the field observed in the two distinct Rindler regions \cite{walls}. The UA
observer can neither access nor influence field modes in the causally
disconnected region and so information is lost about the quantum state, resulting in
detection of a thermal state, a phenomenon known as the Unruh effect
\cite{unruh}.

For an NUA observer consider the coordinates $(T,X)$
\begin{equation}
\label{eq1} w(t+x)=2\sinh(w(T+X)) \ \  w(t-x)=-e^{-w(T-X)}%
\end{equation}
yielding the metric
\begin{equation}
\label{eq2}ds^{2}=\left[  \exp(-2wT)+\exp(2wX)\right]  [dT^{2}-dX^{2}]
\end{equation}
which covers $t-x<0$, referred to as region (I). The remaining half-plane
$t-x>0$ (region (II)) can be covered using  coordinates (\ref{eq1}) after
making the substitutions $X\rightarrow-X$, $T\rightarrow-T$ and $w\rightarrow
-w$. This observer's acceleration
\begin{equation}
\label{eq3}a(T,X_{0})=\left[  e^{-2wT}+ e^{2wX_{0}}\right]^{-3/2}%
w\exp(2wX_{0})
\end{equation}
along $X=X_{0}$ is non-uniform, with $a(T\rightarrow-\infty,X_{0})=0$ and
$a(T\rightarrow+\infty,X_{0})=w\exp(-wX_{0})$. This observer, restricted to
the region $t-x<0$ can causally influence events in $t-x>0$ but cannot be
causally influenced by them. Consequently one expects a loss of information
about the state of a quantum field, with an associated degradation of entanglement.

For both the Rindler and NUA observers entropy no longer quantifies
entanglement because an entangled pure state seen by inertial
observers appears mixed from their frames. However for the NUA
observer this mixed state changes with time, making its description
somewhat problematic. Furthermore, since $\partial/\partial T$ is
not a Killing vector there is no clear way to define a vacuum state.

Our approach is to describe  the field modes on a
given spacelike hypersurface at time $T_0$. On any such hypersurface
instantaneous positive and negative frequencies (with their respective annihilation/creation operators)
can be defined and a solution
to the Klein-Gordon equation in $(T,X)$ coordinates can be related to one in
$(t,x)$ coordinates by a Bogoliubov transformation at $T=T_{0}$.   It is then
possible to bound the entanglement of the mixed state using logarithmic
negativity, which is a full entanglement monotone that bounds distillable
entanglement from above \cite{vidal}. This quantity will depend on $T_{0}$,
and so one can track the change in entanglement as $T_{0}$ is varied. Similar
considerations apply for mutual information \cite{mutual}, which can be used to
quantify the state's total correlations (classical plus quantum). We find that
two modes maximally entangled in an inertial frame become less entangled as a
function of $T_{0}$, with the entanglement approaching a fixed value as
$T_{0}\to\infty$.

The Klein Gordon equation in $(T,X)$ coordinates is
\begin{equation}
\label{eq4}\left[
\partial_{T}^{2}-\partial_{X}^{2}+m^{2}(e^{-2wT}+e^{2wX})\right]  \Phi(X,T)=0
\end{equation}

Separating variables via $\Phi(X,T)=F(T)G(X)$ yields the second
order differential equations
\begin{equation}
\label{eq5}\frac{d^{2}F(T)}{dT^{2}}+(m^{2}\exp(-2wT)+K^{2})F(T)=0
\end{equation}
\begin{equation}
\label{eq6}\frac{d^{2}G(X)}{dX^{2}}-(m^{2}\exp(2wX)-K^{2})G(X)=0
\end{equation}
where  $K^2$ is a constant of separation. Writing $\nu=K/w$,  the solutions
\begin{equation}
\label{eq7}F(T)=A_{\nu}J_{i\nu}(\frac{m}{w}e^{-wT})+B_{\nu}J_{-i\nu}%
(\frac{m}{w}e^{-wT})
\end{equation}
\begin{equation}
\label{eq8}G(X)=C_{\nu}K_{i\nu}^{1}(\frac{m}{w}e^{wX})+D_{\nu}I_{i\nu}%
^{2}(\frac{m}{w}e^{wX})
\end{equation}
can be expressed as linear combinations of Bessel functions $J_{\mu}(Z)$ and McDonald functions $K_{\mu}^{1}(Z)$ and $I_{\mu}^{2}(Z)$.

As $T\to\infty$ the solutions to (\ref{eq5}) asymptote to $F\sim e^{\mp iKT}$,
similar to those for a Rindler observer with time coordinate $\tau=T$, whereas
for $T\to-\infty$ the solutions asymptote to to $F\sim\exp(\mp i\frac{m}%
{w}e^{-wT})$, similar to those for an inertial observer with time coordinate
$wt=-e^{-wT}$. Requiring the solutions to be nonsingular in $X$ we obtain from
(\ref{eq7},\ref{eq8}) the positive frequency solution with the former
asymptotic behaviour is
\begin{equation}
\label{eq9}\Phi_{\nu}^{\mathrm{R}+}(X,T)=(\frac{\nu}{\pi w})^{1/2} J_{i\nu}%
(\tilde{T})K_{i\nu}^{1}(\tilde{X})
\end{equation}
where $\tilde{T}=\frac{m}{w}e^{-wT}$, $\tilde{X}=\frac{m}{w}e^{wX}$.
The solution with the latter asymptotic behaviour is
\begin{equation}
\label{eq10}\Phi_{\nu}^{\mathrm{I}+}(X,T)=\frac{\sqrt{1-e^{-2\pi\nu}}}{2}(\frac{\nu
}{\pi w})^{1/2}H_{i\nu}^{1}(\tilde{T})K_{i\nu}^{1}(\tilde{X})
\end{equation}
as can be determined from the asymptotic behavior of the Bessel functions. The
solutions (\ref{eq9},\ref{eq10}) are orthogonal
\begin{equation}
\label{eq11} \langle\Phi_{\mu}^{\mathrm{R+}}(X,T),\Phi_{\nu
}^{\mathrm{I+}}(X,T)\rangle=(w e^{\pi\nu} \sqrt{e^{2\pi\nu}-1})^{-1}\delta\left(\mu-\nu\right)%
\end{equation}
under the inner product
\begin{equation}
\label{eq12}\langle\Phi,\Psi\rangle=-i\int_{S}dS^{\alpha}
\Phi\overleftrightarrow{{{\partial}}}
_{\alpha}\Psi^{\ast}
\end{equation}
where $dS^{\alpha}$ is the measure  orthogonal to the Cauchy surface $S$ with
timelike unit normal $n^\alpha$.

We proceed in a way analogous to that applied in discussing particle production in expanding universes \cite{Fulling}. Consider now the hypersurface $T=T_{0}$. We consider a subspace of solutions of  Eq. (\ref{eq4}),  possessing instantaneous positive frequency at the hypersurface $T_{0}$.
A general solution at time $T$
\begin{equation}
\label{eq13}\Phi_{\nu}^{+}(X,T)=(c_{\nu+}\left(  T_{0}\right)  J_{i\nu}(\tilde{T})+c_{\nu-}\left(  T_{0}\right)  J_{-i\nu}(\tilde{T}%
))K_{i\nu}(\tilde{X})
\end{equation}
will be a positive frequency solution at $T=T_{0}$ provided
\begin{equation}
\label{eq14}c_{\nu\pm}\left(  T\right)  =\frac{\mp i\nu\pi W^{1/2}}{2K\sinh(\pi\nu
)}(\dot{J}_{\mp i\nu}(\tilde{T})W^{-1}+iJ_{\mp i\nu}(\tilde{T}))
\end{equation}%
where $W=\sqrt{m^{2} e^{-2wT}+K^{2}}$ and the overdot denotes a derivative
with respect to $T$.

A solution for the NUA observer can be expressed in terms of positive and
negative frequency solutions
\begin{equation}
\label{eq16}\Phi_{k}^{\mathrm{M\pm}}(x,t)=\frac{1}{2}(\pi\epsilon)^{-1/2}\exp(\mp
i(\epsilon t-kx))
\end{equation}
of the inertial observer via the Bogoliubov transformations
\begin{equation}
\label{eq17}\alpha_{\nu,k}=\langle\Phi_{\nu}^{\mathrm{+}}(X,T),\Phi
_{k}^{\mathrm{M+}}(x,t)\rangle
\end{equation}%
\begin{equation}
\label{eq18}\beta_{\nu,k}=-\langle\Phi_{\nu}^{\mathrm{+}}(X,T),\Phi
_{k}^{\mathrm{M-}}(x,t)\rangle
\end{equation}
where $\epsilon^2 = k^2+m^2$.
For the positive frequency solution (\ref{eq13}) eqs.
(\ref{eq17},\ref{eq18}) give at $T=T_{0}$
\begin{align}
\alpha_{\nu,k} & = -\mathcal{C}c_{\nu+}(T_0) A_{\nu,k} +
\mathcal{C}c_{\nu-}(T_0) B_{\nu,k}^{\ast
}\label{eq19}\\
\beta_{\nu,k} & = \mathcal{C}c_{\nu+}(T_0) B_{\nu,k}+\mathcal{C}c_{\nu-}(T_0) A_{\nu,k}^{\ast
}\label{eq20}%
\end{align}
where
\begin{equation}
\label{eq21}A_{\nu,k}=-(\frac{\nu}{\epsilon w})^{1/2}\frac{e^{\pi\nu/2}}%
{2\Gamma(1+i\nu)\sinh(\pi\nu)}(\frac{\epsilon -k}{2 w})^{i\nu}%
\end{equation}
${B_{\nu,k}}=-{A_{\nu,k}}e^{-\pi\nu}$ and $\mathcal{C}=\sqrt{\frac{\pi w}{\nu}}$.

\section{Quantification of the entanglement}

Consider now two modes, $k$ and $s$, of a free scalar field in
Minkowski spacetime that are maximally entangled from an inertial perspective
\begin{equation}
\label{eq22}\frac{1}{\sqrt{2}}\left(  \left|  0_{s}\right\rangle
^{\mathcal{M}}\left| 0_{k}\right\rangle ^{\mathcal{M}}+\left|  1_{s}%
\right\rangle ^{\mathcal{M}}\left|  1_{k}\right\rangle ^{\mathcal{M}}\right)
\end{equation}
where $\left|  0_{j}\right\rangle ^{\mathcal{M}}$ and $\left| 1_{j}%
\right\rangle ^{\mathcal{M}}$ are the vacuum and single particle excitation
states of the mode $j$ with respect to an inertial observer in Minkowski space. We assume that Alice can only
detect mode $s$ with her detector and that Vic can only detect mode $k$ with
his.
The Minkowski vacuum state $
|0\rangle^{\mathcal{M}}=\prod_{j}|0_{j}\rangle^{\mathcal{M}}$
is defined as the absence of any particle
excitation in any of the modes.  It can be expressed in terms of the  vacuum $|0\rangle^{\mathcal{A}}$ with respect to the accelerated observer as\cite{Letaw}
\begin{equation}\label{eq23a}
|0\rangle^{\mathcal{M}}=\frac{1}{{}^{\mathcal{M}}\langle 0|0\rangle^{\mathcal{A}}}\exp (-\frac{1}{2}\Sigma_{\nu,k}\tilde{a}_{\nu}V_{ \nu \lambda}\tilde{a}_{\lambda}^{\dagger})|0\rangle^{\mathcal{A}}
\end{equation}
Using  $(\alpha_{\nu,k}^{*})_{I}=(\alpha_{\nu,k})_{II}$ and $(\beta_{\nu,k}^{*})_{I}=(\beta_{\nu,k})_{II}$, where  subindices I and II indicate respectively regions (I) and (II),  we get
\begin{eqnarray}
\label{eq25} V_{\nu ,\lambda}&=& \sum_{l}\beta_{ \nu l}^{*}\alpha_{l \lambda}^{-1}=-\delta(\nu -\lambda) q
\end{eqnarray}
where $q$ is given by
\begin{equation}
q=\frac{e^{-\pi
\nu}c_{\nu+}^{*}(T_0)-c_{\nu-}^{*}(T_0)}{(c_{\nu+}(T_0)+e^{-\pi \nu}c_{\nu-}(T_0))}
\end{equation}
We then rewrite (\ref{eq23a}) in terms of a product of two-mode squeezed states of the NUA vacuum defined on the hypersurface $T=T_{0}$
\begin{equation}
\label{eq24}|0>^{M}=\sqrt{1-|q|^{2}}\sum_{n=0}^{\infty}q^{n}
|n_{k}>_{I}|n_{k}>_{II}%
\end{equation}
where $\left|  n_{k}\right\rangle _{I}$ and $\left|
n_{k}\right\rangle _{II}$ refer to the mode decomposition in regions
(I) and (II), respectively. Each Minkowski mode $j$ has a mode
expansion given by Eq. (\ref{eq24}). We assume that all modes except
for mode $s$ for Alice and $k$ for Vic are in the vacuum. Tracing
over all of these other   modes yields a pure state since each set
of solutions in regions (I) and (II) are orthogonal and so different
modes $j$ and $j^{\prime}$ do not mix on the hypersurface $T=T_{0}$.

Since events in region (II) cannot causally influence those in region (I) we
rewrite Eq.(\ref{eq22}) using (\ref{eq24}), tracing over states in region
(II). This yields a mixed state  between Alice (A) and Vic (V)
\begin{align}
\rho_{AV}  & =\frac{1-|q|^{2}}{2}\sum_{n}q^{2n}\rho_{n},\\
\rho_{n}  &
=|0\,n\rangle\langle0\,n|+\sqrt{1-|q|^{2}}\sqrt{n+1}|0\,n\rangle
\langle1\,n+1|\nonumber\\
& +\sqrt{1-|q|^{2}}\sqrt{n+1}|1\,n+1\rangle\langle0\,n| \nonumber \\
&+(1-|q|^{2})\left( n+1\right)|1\,n+1\rangle\langle1\,n+1|\nonumber
\end{align}
whose elements are functions of $T_0$ via the parameter $q$, where $|n\,m\rangle=|n_{s}\rangle^{\mathcal{M}}|m_{k}\rangle_{I}$.

\begin{figure}
\includegraphics[width=7cm]{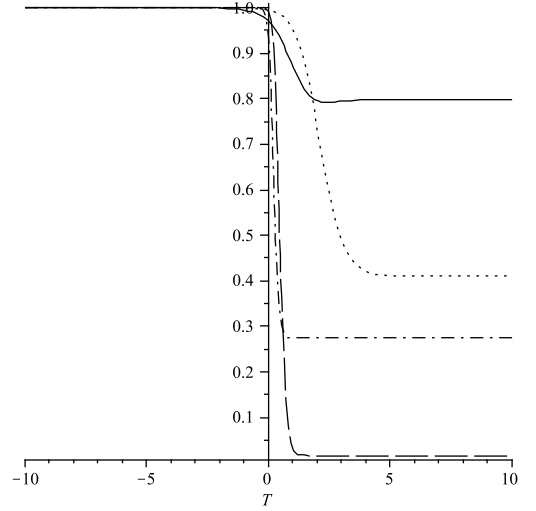}
\caption{Logarithmic negativity  plotted against $T_0$, with $\nu=K/w$,  
for various values of $K$ and $w$.  a)  The dot line corresponds to $K=0.1$, $w=1$,  b) the solid line corresponds
to $K=0.3$, $w=1$, c) the dash line corresponds to $K=0.1$, $w=5$, and d) the dash-dot line corresponds to $K=0.3$, $w=5$
 \label{fig1}}
\end{figure}

\begin{figure}
\includegraphics[width=7cm]{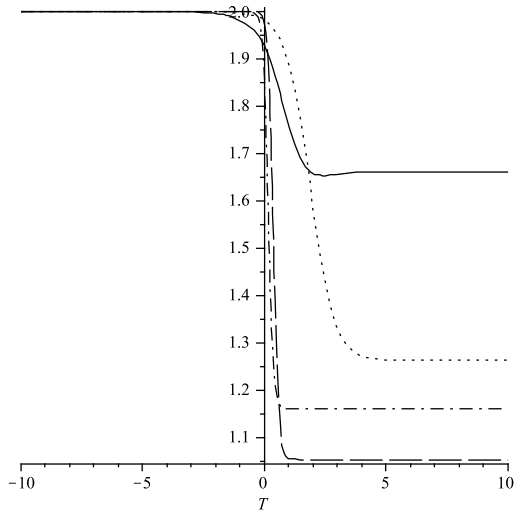}
\caption{Mutual information plotted against $T_0$, with $\nu=K/w$,  
for various values of $K$ and $w$.  a)  The dot line corresponds to $K=0.1$, $w=1$,  b) the solid line corresponds
to $K=0.3$, $w=1$, c) the dash line corresponds to $K=0.1$, $w=5$, and d) the dash-dot line corresponds to $K=0.3$, $w=5$
\label{fig2}}
\end{figure}

From here the calculation proceeds in a manner similar to that for the Rindler case \cite{ivette}.
If at least one eigenvalue of the partial transpose of $\rho$ is negative, then the density matrix is
entangled; but a state with positive partial transpose can still
be entangled. This criterion is sufficient for determining
entanglement \cite{peres}, whose type in this case is called bound or
non-distillable entanglement \cite{vidal}. We obtain the partial
transpose by interchanging Alice's qubits obtaining
\begin{equation*}
\lambda _{\pm }^{n}=\frac{1}{4}{|q|^{2n}(1-|q|^{2})}
\left[ \left( \frac{n |q|^{2}}{\sqrt{1-|q|^{2}}}+|q|^{2}\right) \pm
\sqrt{Z_{n}}\right] ,
\end{equation*}
for the eigenvalues in the $\left(n,n+1\right) $ block,  where
\begin{equation}
Z_{n}=\left( \frac{n |q|^{2}}{\sqrt{1-|q|^{2}}} + |q|^{2} \right)
^{2}+ 4 (1-|q|^{2}) .  \notag
\end{equation}
One eigenvalue is always negative since $|q|<1$ for any value of $T_0$ and so the state is always entangled.
Summing over all the negative eigenvalues yields the
logarithmic negativity \cite{vidal},  defined as
$ N(\rho )=\log _{2}||\rho ^{T}||_{1}$
where $||\rho ^{T}||_{1}$ is
the trace-norm of the density matrix $\rho$. The  quantity $N(\rho )$ bounds the distillable entanglement (the amount of `almost-pure state entanglement' that can be asymptotically distilled) contained in
$\rho$.  We find
\begin{equation}
N(\rho _{AV})=\log _{2}\left(\frac{1-|q|^2}{2}+\sum_{n=0}^{\infty }\frac{|q|^{2n}}{4}{(1-|q|^{2})}
\sqrt{Z_{n}}\right)
\end{equation}
Fig. \ref{fig1} shows that logarithmic negativity decreases with
increasing time, its slope more pronounced close to $T_0=0$ where
the change in acceleration is maximal. For vanishing acceleration
($q\rightarrow 0$ as $T\rightarrow -\infty$), $N(\rho _{AV})\to 1$,
and as  $T_0\rightarrow +\infty$, it approaches the values obtained
for the UA case \cite{ivette},  with different values of $K=\nu w$
yielding different asymptotic values of $N(\rho _{AV})$.   

Hence entanglement degradation increases as a function of time. This
behaviour could in principle be empirically determined by Alice from
an ensemble of experiments with different NUA observers each with
the  same value of $w$, making measurements for different choices of
$T_0$ that are classically communicated to Alice.

To estimate the total amount of correlation in the state we compute
the mutual information, defined as
$I(\rho_{AV})=S(\rho _{A})+S(\rho _{V})-S(\rho _{AV})$ where
$S(\rho )=-\text{Tr}(\rho \log _{2}(\rho ))$ is the entropy of the
density matrix $\rho $. We obtain Alice's density matrix $\rho _{A} = \frac{1}{2}I$ by tracing over Vic's
states and Vic's density matrix  by tracing over Alice's states.  The resultant entropies can be straightforwardly calculated along with the entropy of the joint state, yielding
\begin{eqnarray*}
I  &=&1-\frac{1}{2}\log _{2}\left( |q|^2\right) -\frac{1-|q|^2}{2}\sum_{n=0}^{\infty}|q|^{2n}\mathcal{D}_{n}, \\
\mathcal{D}_{n} &=&(1+\frac{n(1-|q|^{2})}{|q|^2})\log _{2}\left( {1+\frac{n(1-|q|^2)}{|q|^2}}\right) \\
&-&(1+(n+1)(1-|q|^2)\log _{2}\left( 1+(n+1)(1-|q|^2) \right),
\end{eqnarray*}%
for the mutual information, which we plot in Fig. \ref{fig2} as a function of $T_0$. For large negative values of $T_0$ the acceleration vanishes
and the the mutual information is $2$ as expected. As the acceleration
increases, the mutual information decreases reaching, as $T_0\rightarrow +\infty$,  a constant
value larger than unity. 

The logarithmic negativity as well as the mutual information approach for large values of $T$ a constant  asymptotic value that 
depends on $\nu=K/w$. Fig. \ref{fig1} and Fig. \ref{fig2} show that, as expected from the Rindler case \cite{ivette},  for a given value of $K$, the asymptotae of  $N$ and $I$ decrease as
$w$ increases. We also see that for a given vale of $w$, the asymptotic values of $N$ and $I$ increase monotonically with $K$.  

\section{Concluding remarks}

Our method for tracking the time-dependence of measures of quantum information
via a sequence of measurements on hypersurfaces where positive/negative frequencies can be defined
 is quite general and can be applied to different kinds of motions and fields beyond the example we consider here.    For a situation in which both observers begin freely falling into a black hole with
one observer increasing acceleration to avoid this fate,  the distillable entanglement degrades to a finite value.  The entanglement
degradation is due to the increase of entanglement with the modes in the region causally undetectable
by the NUA observer. In curved spacetime, we expect in general that entanglement is a time-dependent as well as an observer-dependent concept.

This work was supported in part by the Natural Sciences \& Engineering Research Council of Canada.


\end{document}